\documentclass[%
 aip,
 amsmath,amssymb,
 reprint,
]{revtex4-1}

\usepackage{graphicx}
\usepackage{dcolumn}
\usepackage{bm}
\usepackage[utf8]{inputenc}
\usepackage[T1]{fontenc}
\usepackage{etoolbox}
\usepackage[english]{babel}
\usepackage{xcolor}
\usepackage{natbib}
\draft

\begin{document}

\title{Magnetic shocks generated by two counterstreaming plasma jets induced by an intense femtosecond laser pulse}

\author{R. S. Zemskov}
\affiliation{A.V. Gaponov-Grekhov Institute of Applied Physics of the Russian Academy of Sciences (IAP RAS), 46 Ulyanov st., Nizhny Novgorod 603950, Russia}

\author{A. V. Kotov}
\affiliation{A.V. Gaponov-Grekhov Institute of Applied Physics of the Russian Academy of Sciences (IAP RAS), 46 Ulyanov st., Nizhny Novgorod 603950, Russia}

\author{S. E. Perevalov}
\affiliation{A.V. Gaponov-Grekhov Institute of Applied Physics of the Russian Academy of Sciences (IAP RAS), 46 Ulyanov st., Nizhny Novgorod 603950, Russia}
 
  \author{A.A. Murzanev}
\affiliation{A.V. Gaponov-Grekhov Institute of Applied Physics of the Russian Academy of Sciences (IAP RAS), 46 Ulyanov st., Nizhny Novgorod 603950, Russia}

\author{A. I. Korytin}
\affiliation{A.V. Gaponov-Grekhov Institute of Applied Physics of the Russian Academy of Sciences (IAP RAS), 46 Ulyanov st., Nizhny Novgorod 603950, Russia}

\author{P. A. Chuvakin}
\affiliation{A.V. Gaponov-Grekhov Institute of Applied Physics of the Russian Academy of Sciences (IAP RAS), 46 Ulyanov st., Nizhny Novgorod 603950, Russia}

\author{A.V. Romashkin}
\affiliation{A.V. Gaponov-Grekhov Institute of Applied Physics of the Russian Academy of Sciences (IAP RAS), 46 Ulyanov st., Nizhny Novgorod 603950, Russia}

 \author{V.N. Ginzburg}
\affiliation{A.V. Gaponov-Grekhov Institute of Applied Physics of the Russian Academy of Sciences (IAP RAS), 46 Ulyanov st., Nizhny Novgorod 603950, Russia}

 \author{A.A. Kochetkov}
\affiliation{A.V. Gaponov-Grekhov Institute of Applied Physics of the Russian Academy of Sciences (IAP RAS), 46 Ulyanov st., Nizhny Novgorod 603950, Russia}

 \author{S.E. Stukachev}
\affiliation{A.V. Gaponov-Grekhov Institute of Applied Physics of the Russian Academy of Sciences (IAP RAS), 46 Ulyanov st., Nizhny Novgorod 603950, Russia}

 \author{I.V. Yakovlev}
\affiliation{A.V. Gaponov-Grekhov Institute of Applied Physics of the Russian Academy of Sciences (IAP RAS), 46 Ulyanov st., Nizhny Novgorod 603950, Russia}

 \author{I.A. Shaikin}
\affiliation{A.V. Gaponov-Grekhov Institute of Applied Physics of the Russian Academy of Sciences (IAP RAS), 46 Ulyanov st., Nizhny Novgorod 603950, Russia}

 \author{A.A. Kuzmin}
\affiliation{A.V. Gaponov-Grekhov Institute of Applied Physics of the Russian Academy of Sciences (IAP RAS), 46 Ulyanov st., Nizhny Novgorod 603950, Russia}

 \author{K.F. Burdonov}
\affiliation{A.V. Gaponov-Grekhov Institute of Applied Physics of the Russian Academy of Sciences (IAP RAS), 46 Ulyanov st., Nizhny Novgorod 603950, Russia}

\author{A. A. Soloviev}
\affiliation{A.V. Gaponov-Grekhov Institute of Applied Physics of the Russian Academy of Sciences (IAP RAS), 46 Ulyanov st., Nizhny Novgorod 603950, Russia}

 \author{A.A. Shaykin}
\affiliation{A.V. Gaponov-Grekhov Institute of Applied Physics of the Russian Academy of Sciences (IAP RAS), 46 Ulyanov st., Nizhny Novgorod 603950, Russia}

 \author{A. N. Stepanov}
\affiliation{A.V. Gaponov-Grekhov Institute of Applied Physics of the Russian Academy of Sciences (IAP RAS), 46 Ulyanov st., Nizhny Novgorod 603950, Russia}

\author{M. V. Barkov}
\email[]{barkov@inasan.ru}
\affiliation{INASAN - Institute of Astronomy, Russian Academy of Sciences, 119017, Moscow, Russia}
\affiliation{Joint Institute for Nuclear Research, 141980, Dubna, Russia}

\author{E. A. Khazanov}
\affiliation{A.V. Gaponov-Grekhov Institute of Applied Physics of the Russian Academy of Sciences (IAP RAS), 46 Ulyanov st., Nizhny Novgorod 603950, Russia}

\author{M. V. Starodubtsev}
\email[]{mstar@ipfran.ru}
\affiliation{A.V. Gaponov-Grekhov Institute of Applied Physics of the Russian Academy of Sciences (IAP RAS), 46 Ulyanov st., Nizhny Novgorod 603950, Russia}

\date{\today}

\begin{abstract}
This work extends the investigation of magnetic control of counterstreaming laser-produced plasma jets reported in Zemskov et al. (2026) \cite{zemskov2026magnetic}, focusing on the experiments performed in the presence of a uniform 14 T magnetic field oriented transverse to the jet propagation. When generating plasma flows using 60--80 fs, 10--15 J pulses at the PEARL facility, we observe the formation of narrow, stationary layers with strong magnetic-field amplification and sharp plasma-density gradients in the interaction region. We interpret these features as the result of compressing the externally applied transverse field by the converging plasma ram pressure. The external magnetic field is advected and compressed until its magnetic pressure balances the decreasing ram pressure of the jets. Time-resolved interferometry, shadowgraphy, and Faraday polarimetry, combined with hybrid PIC--fluid simulations, reproduce the formation of a high-field barrier that decelerates and laterally redirects plasma flows while producing multiple density jumps. We discuss the astrophysical implications: the compression of an initially weak ambient field by converging flows provides an efficient, non-turbulent channel for magnetic-field amplification that can render interstellar magnetic fields dynamically important — a condition essential for diffusive shock acceleration in supernova remnants. Our experiments therefore establish a laboratory platform bridging externally magnetized and self-magnetized regimes and offer a complementary route to magnetic-field amplification in high-energy astrophysical shocks.
\end{abstract}

\maketitle

\section{Introduction}

Magnetic-field amplification (MFA) at astrophysical shocks is a fundamental problem in high-energy astrophysics. 
Observations of young supernova remnants (SNR), including thin synchrotron X-ray rims and rapid variability, imply downstream magnetic fields $(100-1000 \;\mu G)$ far in excess of typical interstellar values $(1-3\; \mu G)$ \cite{uchiyama2007extremely}. The bright filaments in X-ray or radio range from fast moving neutron stars also show the presence of a strong magnetic field in the intestellar media (ISM) at the level $30-100\;\mu G$ \cite{2019MNRAS.485.2041B,2027JHEAp..5500710N}.

Such amplified fields in SNR are widely regarded as essential for efficient ion acceleration via diffusive shock acceleration (DSA) \cite{caprioli2014simulations, giacalone2007magnetic, bell2004turbulent}. In the case of young SNR, simple shock compression alone cannot account for the inferred magnetic field amplification (see Ref.\cite{2027JHEAp..5500710N}). Modern theory and large-scale simulations have, therefore, proposed several complementary mechanisms: cosmic-ray driven instabilities (resonant and non-resonant Bell modes) that grow in the shock precursor \cite{caprioli2014simulations, bell2004turbulent}; shock--turbulence interactions that produce downstream vortical motions and turbulent dynamo action \cite{giacalone2007magnetic}; and hydrodynamic instabilities such as the Richtmyer--Meshkov instability (RMI) that stretch and fold frozen-in magnetic fields at corrugated interfaces \cite{kuramitsu2016model}.

Unlike young SNR MFA, the filaments in ISM, which were discovered in the X-ray \cite{2008AIPC..983..171K} and in the radio \cite{1985AJ.....90.2511M}, are long thin structure with length up to tenths of pc. These structures could be formed in several ways. In the Galactic center the most feasible scenario is stretching of magnetic field lines in outflow from the central gas disc \cite{2019MNRAS.489L..28B}. In the case of the X-ray filaments in ISM, the most probable scenario is compression on the shock of the old SNR \cite{2027JHEAp..5500710N} or interaction of multiple shocks.

Direct astrophysical validation of MFA mechanisms is difficult because the small-scale plasma dynamics involved is not directly observable in SNRs and other astrophysical objects. Laboratory high-energy-density platforms offer a complementary route by reproducing scaled dimensionless parameters in controlled configurations, where specific mechanisms can be isolated and tested experimentally \cite{Ryutov_2000, Ryutov_1999}. In recent years, laser-plasma experiments have demonstrated key processes relevant to MFA: Biermann battery field generation and advection, Weibel-mediated filamentation, collisionless shock formation, and RMI-driven field amplification \cite{courtois2004experiment, fiksel2014magnetic, kugland2013visualizing, kuramitsu2016model, zemskov2026magnetic, zemskov2025non}.

In this work we explore a distinct and experimentally accessible amplification channel: the advection and compression of an externally applied transverse magnetic field by converging, counterstreaming plasma jets. When the magnetic Reynolds number is large, the ambient field is effectively frozen into the plasma on the interaction timescale and cannot diffuse deeply into the incoming flows. Converging plasmas with high ram pressure, therefore, compress and amplify the transverse field until the amplified magnetic pressure balances the plasma pressure, producing a narrow, high-field magnetic layer bordered by sharp local plasma density gradients. This compression-driven amplification operates on timescales much shorter than those of turbulent magnetic field amplification, offering a complementary pathway to amplify weak ambient fields to dynamically important strengths.

We present experiments performed at the PEARL facility, in which two femtosecond-driven plasma jets collide in the presence of a uniform 14 T magnetic field applied transverse to the jet axis. Time-resolved interferometry, shadowgraphy, and Faraday polarimetry, supported by hybrid PIC--fluid simulations, reveal the formation of compressed magnetic layers, multiple sharp density jumps, and lateral redirection of plasma flows. We compare these findings with the previous laboratory and theoretical MFA studies \cite{courtois2004experiment, kuramitsu2016model, giacalone2007magnetic, caprioli2014simulations}, and discuss the implications for astrophysical shocks, the filaments in
ISM and DSA, where converting a weak interstellar field into one that can magnetize ions is a prerequisite for efficient particle acceleration.

The paper is organized as follows. Section \ref{sec:Methods} describes the setup and experimental scheme. Section \ref{sec:Res} presents the results of the experiment. Section \ref{sec:Sym} provides the results of numerical modeling using the hybrid code AKA52. Section \ref{sec:Discus} presents a discussion of the obtained results and the interpretation of the numerical computations.
 

\section{\label{sec:Methods} Methods}

The experiments were carried out at the PEARL petawatt laser facility \cite{lozhkarev2007compact, ginzburg202111}. The main pump pulse had an energy of 10--15 J, a duration of 60--80 fs, and a wavelength of 910 nm. It was focused by an $F/8$ spherical mirror into a spot of roughly 100 $\mu$m diameter, producing a peak intensity of approximately $2 \times 10^{18}$ W/cm$^{2}$ on the target surface. The intensity of the pre-pulse was kept below $10^{-7}$ of the main pulse. Further details on the facility and diagnostic approach are available in Ref. \cite{zemskov2026magnetic} and in the review \cite{soloviev2024research}.

\begin{figure}
 \includegraphics[width=8 cm]{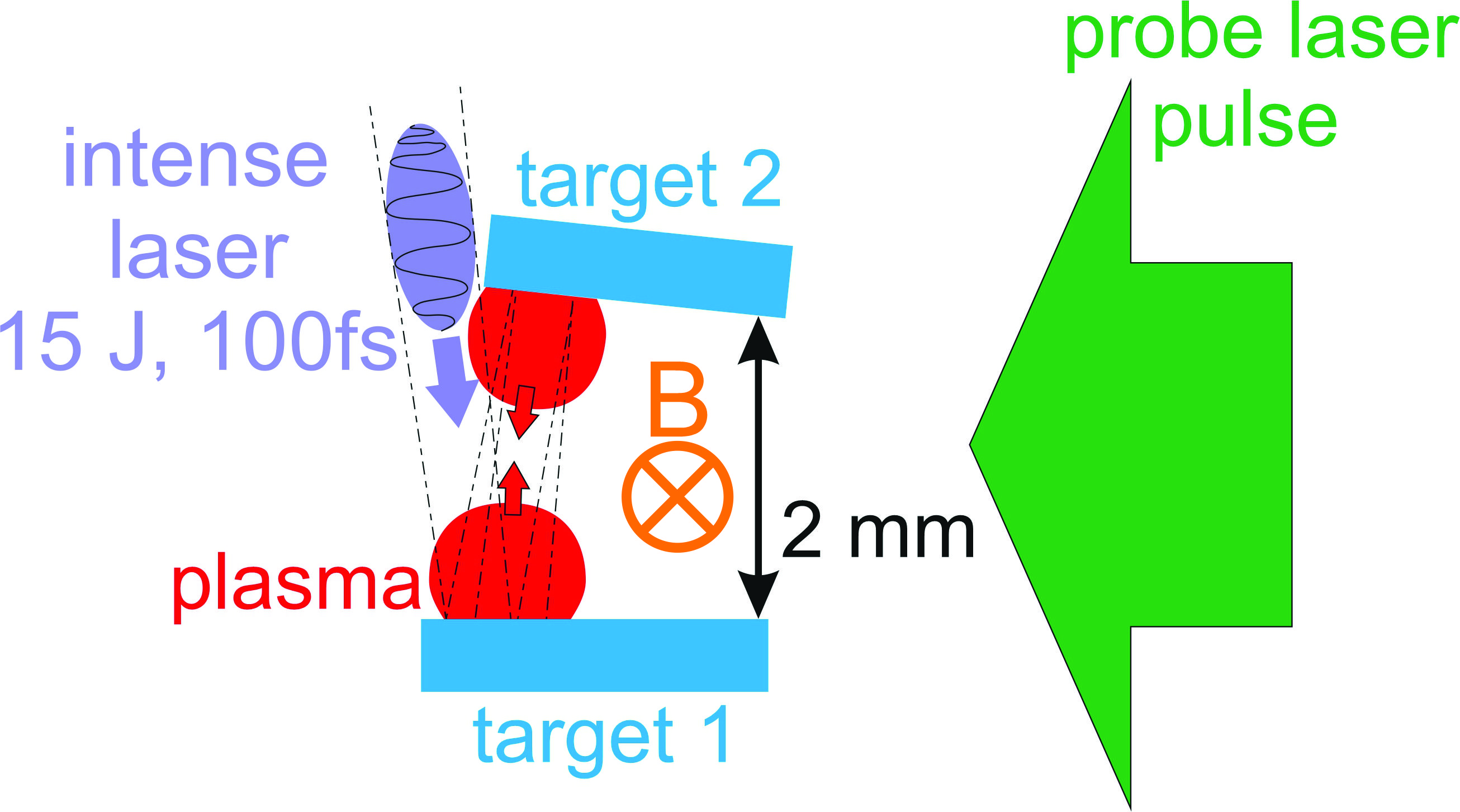}
 \caption{\label{fig:Scheme} Experimental arrangement for generating counterstreaming plasma jets. The pump pulse irradiates the first PMMA target and is reflected toward the second PMMA target, creating two plasma jets that meet in the central gap. Orthogonal optical probes are used for shadowgraphy, interferometry, and Faraday polarimetry.}
\end{figure}

A single pump pulse was used to generate two counterstreaming plasma jets. The primary polymethyl methacrylate (PMMA) target was irradiated at a small incidence angle of about 5$^{\circ}$ relative to the target normal, with the laser polarization in the target plane (s-polarization) (Fig. \ref{fig:Scheme}). A second PMMA target was positioned 2 mm from the first and aligned so that the reflected pump beam struck its surface at approximately the same small angle. Laser absorption measurements and previous studies of similar relativistic-intensity interactions \cite{price1995absorption, ping2008absorption} indicate that roughly 20--30\% of the pump energy is absorbed by the first target, while the remainder is reflected and scattered toward the second target. The long Rayleigh length ($\sim$10 mm) ensured similar spot sizes on both targets despite the 2 mm separation. Under our conditions, this implies that the second target received a pulse energy of about 8 J, producing a plasma jet with parameters comparable to those of the first. This setup produced two quasi-opposite plasma jets propagating along the normals of the two targets.

In the absence of the field, the experiment reproduced the configuration reported in Ref. \cite{zemskov2026magnetic}, where self-generated toroidal fields dominated the interaction. The main experimental modification in the present study is the addition of an externally imposed field. A uniform external magnetic field of 14 T was applied along the plane of the targets, transverse to the initial jet propagation direction. The applied magnetic field was produced by a pulsed Helmholtz coil integrated into the target chamber. Its uniformity scale $\gtrsim$ 1 cm across the interaction region was sufficient to ensure that the jet propagation initially occurred in a homogeneous transverse field.

\begin{table*}
\caption{The values of the initial parameters for each of the two flows.} 
\begin{center}
\begin{tabular}{c|c|c|c|c|c|c|c|c}
\hline 
\hline
\multicolumn{1}{c|}{Parameter} & \multicolumn{1}{c|}{Electron density} & \multicolumn{1}{c|}{Charge state} & \multicolumn{1}{c|}{Mass number}  & \multicolumn{1}{c|}{Flow velocity}  & \multicolumn{1}{c|}{Ion temp.} & \multicolumn{1}{c|}{Electron temp.} & \multicolumn{1}{c|}{Spatial scale} & \multicolumn{1}{c}{Magn. field} 
\tabularnewline
& \multicolumn{1}{c|}{$n_e$, cm$^{-3}$} & \multicolumn{1}{c|}{Z} & \multicolumn{1}{c|}{A}  & \multicolumn{1}{c|}{V, km/s}  & \multicolumn{1}{c|}{T$_{i}$, eV} & \multicolumn{1}{c|}{T$_{e}$, $eV$} & \multicolumn{1}{c|}{L, cm}   & \multicolumn{1}{c}{B, T}
\tabularnewline
\hline 

\multicolumn{1}{c|}{Value} & 5 $\times 10^{18}$ & 2 & 7 & 100 & 50 & 50 & 0.1 & 14
\tabularnewline
\hline

\end{tabular}
\end{center}
\label{Table_in}
\end{table*}

The diagnostics suite comprised femtosecond optical probing from an attenuated replica of the main pulse. The probe pulse had an energy near 1 $\mu$J and a duration of 60--80 fs. This permitted near-instantaneous snapshots of the evolving plasma. Electron density maps were obtained by Mach--Zehnder interferometry, while Faraday polarimetry was used to detect magnetic-field-induced rotation of the probe polarization \cite{swadling2014diagnosing}. Shadowgraphy provided complementary morphological images of the plasma structures. Probe beams were delayed from 0.5 ns up to a few tens ns after the pump pulse and were directed along two orthogonal lines of sight in order to capture the three-dimensional shape of the interaction region.

Further details about the experiment, including the exact construction of the vacuum chamber, diagnostics setup and the magnetic field generation system, can be found in Ref. \cite{zemskov2026magnetic} and the review \cite{soloviev2024research}. In this paper we focus on the central difference from the study in Ref. \cite{zemskov2026magnetic}: the jets are produced in the presence of an externally imposed transverse magnetic field, while the underlying pulse, target, and diagnostic techniques remain essentially the same.

\begin{table}
\caption{Calculated values of some parameters.} 
\begin{center}
\begin{tabular}{c|c}

\hline
\hline 
\multicolumn{1}{c|}{$\rho$ $[g.cm^{-3}]$} & $5.9\times10^{-5}$  \tabularnewline
\multicolumn{1}{c|}{C$_{S}$ $[km.s^{-1}]$}  & $58$   \tabularnewline
\multicolumn{1}{c|}{V$_{A}$ $[km.s^{-1}]$}  & $110$  \tabularnewline
\multicolumn{1}{c|}{l$_{ee}$ $[cm]$} & $1.25\times10^{-4}$  \tabularnewline
\multicolumn{1}{c|}{$\tau_{col\:ee}$ $[ns]$}  & $1.65\times10^{-3}$  \tabularnewline
\multicolumn{1}{c|}{R$_{Le}$ $[cm]$}  & $5.6\times10^{-5}$  \tabularnewline
\multicolumn{1}{c|}{f$_{ce}$ $[Hz]$} & $8.4\times10^{11}$ \tabularnewline

\multicolumn{1}{c|}{l$_{ei}$ $[cm]$} & $5\times10^{-6}$ \tabularnewline
\multicolumn{1}{c|}{$\tau_{col\:ei}$ $[ns]$}  & $1.9\times10^{-3}$ \tabularnewline

\multicolumn{1}{c|}{l$_{ii}$ $[cm]$} & $8.7\times10^{-5}$  \tabularnewline
\multicolumn{1}{c|}{$\tau_{col\:ii}$ $[ns]$} & $3.3\times 10^{-3}$  \tabularnewline

\multicolumn{1}{c|}{l$_{i}$ (directed) $[cm]$} & $5\times10^{-2}$  \tabularnewline

\multicolumn{1}{c|}{R$_{Li}$  (directed) $[cm]$} & $1.2\times10^{-2}$ \tabularnewline
\multicolumn{1}{c|}{f$_{ci}$ $[Hz]$} & $1.3\times10^{8}$  \tabularnewline
\multicolumn{1}{c|}{$\tau_{ff}$ $[ns]$} & $9\times10^{3}$  \tabularnewline
\multicolumn{1}{c|}{M} & $1.7$  \tabularnewline
\multicolumn{1}{c|}{M$_{A}$} & $0.9$  \tabularnewline
\multicolumn{1}{c|}{$\eta$ $[cm^2/s]$} & $3.4 \times 10^4$  \tabularnewline
\multicolumn{1}{c|}{Re$_{M}$} & $30$  \tabularnewline
\multicolumn{1}{c|}{Re} & $20$  \tabularnewline
\multicolumn{1}{c|}{Eu} & $2.2$  \tabularnewline
\multicolumn{1}{c|}{$\beta$} & $3 \times 10^{-2}$  \tabularnewline
\hline 

\end{tabular}
\end{center}
$^i$ The following symbols are used: $\rho$ — mass density, $C_{S}$ — sound speed, $V_{A}$ — Alfvén speed $(V_{A} = B/\sqrt{4\pi n_{i} m_{i}}$, where $n_{i}$ is the ion density and $m_{i}$ is the ion mass), $l_{ee}$ — mean free path for e-e collisions, $\tau_{col\:ee}$ — time between electron collisions, $R_{Le}$ —  electron Larmor radius, $f_{ce}$ — electron gyrofrequency, $l_{ii}$ — mean free path for i-i ion collisions, $\tau_{col\:ii}$ — ion collision time, $R_{Li}$ — ion Larmor radius, $f_{ci}$ — ion gyrofrequency, $\tau_{ff}$ —  bremsstrahlung cooling time, $M$ — Mach number, $M_{A}$ — Alfvén Mach number $(M_{A} = V/V_{A})$, Re$_{M}$ — magnetic Reynolds number $(\text{Re}_{M} = LV/\eta)$, $\eta$ — magnetic diffusivity, Re — Reynolds number $(\text{Re} = LV/\nu$, where $\nu$ is the kinematic viscosity), Eu = $\sqrt{\frac{\rho V^2}{p}}$ — Euler number, and $\beta = \frac{p}{\frac{B^2}{8\pi}}$ — plasma beta.
\label{Table_out}
\end{table}

\subsection{\label{subsec:OneFlow} Description of a single laser plasma flow}

\begin{figure}
    \centering
    \includegraphics[width=9 cm]{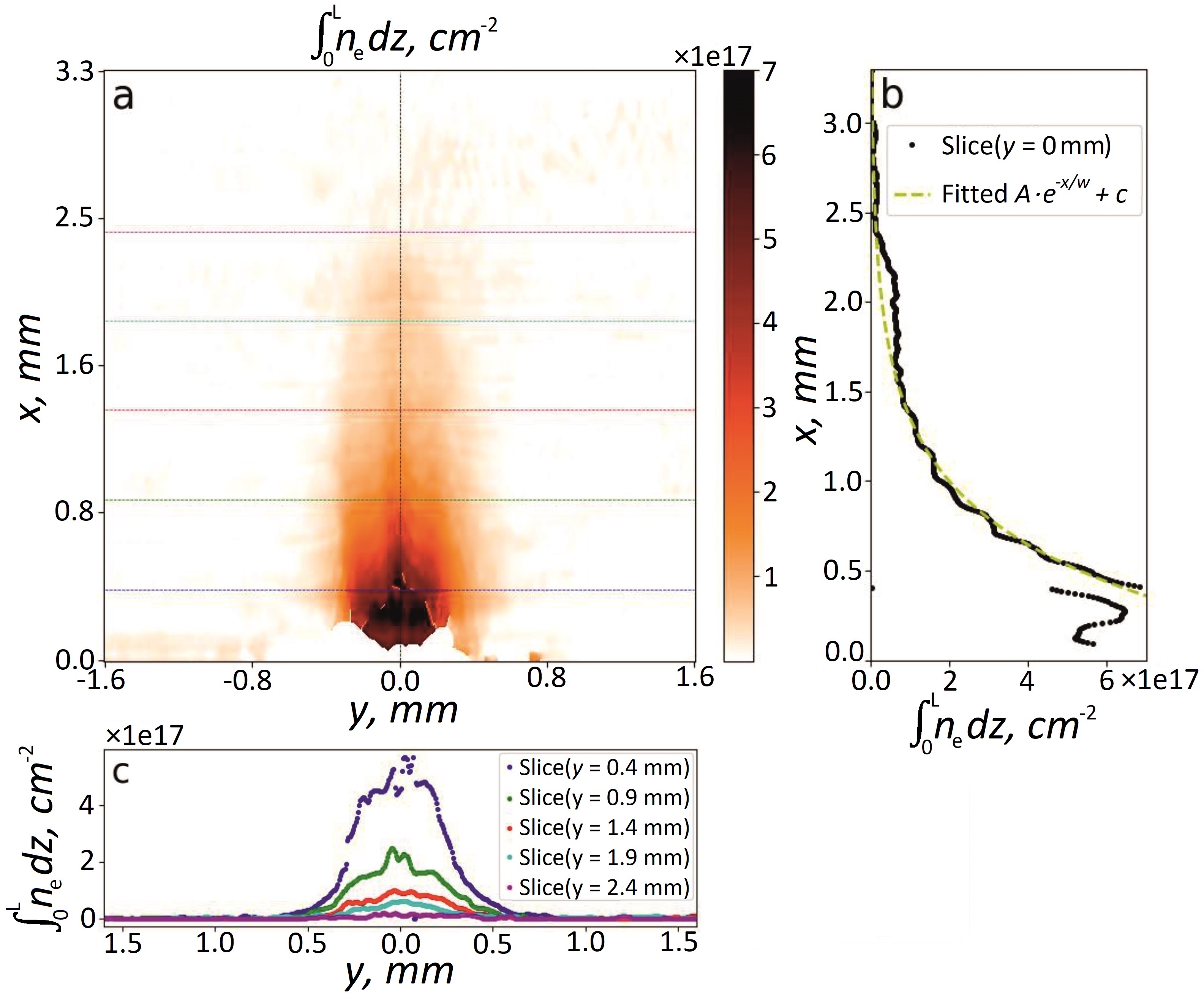}
    \caption{\label{fig:oneflow} 
    Characterization of a single plasma jet 8 ns after irradiation of a single target at $2\times 10^{18}$ W$\cdot$cm$^{-2}$, in the absence of external magnetic field. 
    (a) Two-dimensional map of the integrated plasma density along the line of sight, reconstructed from interferometry. 
    (b) Lineout of the plasma density along the target normal (x-axis), exhibiting an exponential decay with scale length $w \approx 0.5$ mm. 
    (c) Transverse density profiles parallel to the target surface, highlighting the highly collimated geometry of the flow. The figure is adapted from our earlier study \cite{zemskov2026magnetic}.
    }
\end{figure}

Before examining the complex interaction of two counterstreaming jets in an external magnetic field, it is useful to recall the basic properties of a single flow generated under the same laser conditions but without applied magnetic fields, as detailed in our previous work \cite{zemskov2026magnetic}.
In the first few picoseconds after the femtosecond laser irradiation, plasma electrons reach temperatures of several hundred keV \cite{antici2013modeling, dubois2014target, soloviev2017experimental}. 
As the plasma expands into vacuum, it rapidly cools and isotropizes, reaching a characteristic temperature $T_e \sim T_i \sim 50$ eV after a few nanoseconds. 
A typical snapshot of the flow, obtained with femtosecond optical interferometry 8 ns after the drive pulse, is shown in Fig.~\ref{fig:oneflow} (see Ref.~\cite{zemskov2026magnetic} for reconstruction details). 
The plasma forms a highly collimated jet with a divergence angle of no more than a few degrees, in stark contrast to the quasi-spherical expansion typical of nanosecond laser-produced plasmas \cite{Zemskov2025FsNs, zemskov2024laboratory}.

The longitudinal density profile along the target normal (Fig.~\ref{fig:oneflow}b) is well described by an exponential decay $N_e \propto e^{-x/w}$, with $w = 0.5$ mm, consistent with the scaling $N_e \propto e^{-x/(c_s t)}$ for an ion sound speed $c_s \approx 60$ km/s \cite{samir1983expansion}. 
The front velocity of the flow is estimated to be $\sim 100$ km/s. 
For quantitative estimates, the plasma is modeled as a mixture of H$^+$, C$^+$, and O$^+$ ions with an average charge state $\langle Z \rangle \approx 2$ and mass number $\langle A \rangle \approx 7$, yielding a density $n_e \approx 2.5\times 10^{18}$ cm$^{-3}$ at a distance of 1 mm from the target.
The electron mean free path (see Table \ref{Table_out}) does not exceed the characteristic scale of the plasma, confirming that the electrons are in thermodynamic equilibrium, while the mean free path for ion--ion collisions between the counterstreaming flows is comparable to the interaction length.

\section{\label{sec:Res}Results}

\begin{figure*}
    \centering
    \includegraphics[width=16 cm]{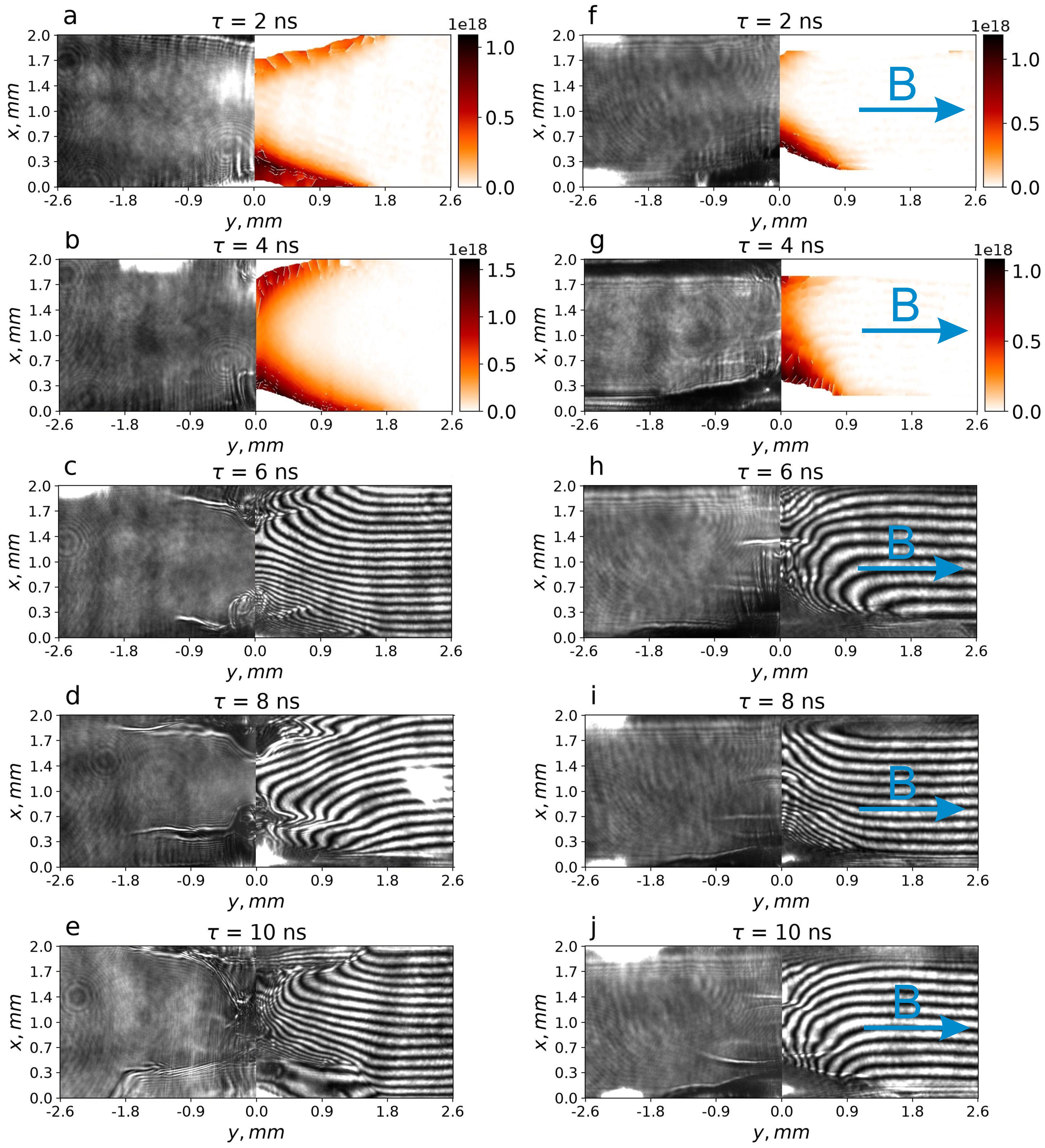}
    \caption{\label{fig:interf2flows} Interferometric snapshots of the interacting jets. The left-hand panels show the reference case from Ref.~\cite{zemskov2026magnetic} without externally applied magnetic field, while the right-hand panels show the corresponding measurements in the case with 14 T external magnetic field.}
\end{figure*}

The experimental evidence is summarized in Fig.~\ref{fig:interf2flows}, which shows the interaction sequence of counter-propagating plasma flows from 2 ns to 10 ns in steps of 2 ns. Each panel is split into two halves along the symmetry axis of the plasma. In the upper panels Fig.~\ref{fig:interf2flows} (a, b, f, g), the left half shows a shadowgraph image and the right half shows the reconstructed linear density of the plasma. In the remaining panels, the left half shows a shadowgraph image and the right half shows an interferometric image. The left-hand panels reproduce the baseline case without an externally applied magnetic field, which was previously discussed in Ref.~\cite{zemskov2026magnetic}; the right-hand panels show the new configuration with an applied magnetic field of 14 T.

In the absence of an external magnetic field, the two plasma jets are initially almost independent (Fig.~\ref{fig:interf2flows} (a)) but,  once their fronts overlap, they rapidly begin to influence each other (Fig.~\ref{fig:interf2flows} (b)). The density maps show that the interaction region broadens substantially and that the central area becomes comparatively depleted, while the plasma remains concentrated in extended fronts that propagate away from the targets (Fig.~\ref{fig:interf2flows} (c, d).

At later times, the morphology becomes more structured (Fig.~\ref{fig:interf2flows} (d, e) . The plasma-density distributions reveal the formation of two compression regions separated by a lower-density gap (Fig.~\ref{fig:interf2flows} (d, e)), and the shadowgraphy images show sharp fronts and a well-defined interaction layer that is no longer consistent with simple interpenetration of the jets (Fig.~\ref{fig:interf2flows} (d, e)). Interferometric measurements confirm the same trend: the plasma accumulates in compressed sideways layers, while the flows continue to interpenetrate along the central axis (Fig.~\ref{fig:interf2flows}). These observations are consistent with the stagnation and lateral redirection of the counterstreaming flows, as reported in Ref.~\cite{zemskov2026magnetic}, where the effect was interpreted as the result of compression of self-generated magnetic field in the interaction zone.

In striking contrast, the magnetized configuration (right-hand panels) exhibits distinctly different behavior at later times. The early stages at 2-4~ns (Fig.~\ref{fig:interf2flows}(f, g)) are similar to the case without external magnetic field, but significant differences emerge after 4~ns. A sharp density discontinuity becomes prominent in the central interaction region in Fig.~\ref{fig:interf2flows} (h), which is absent in the corresponding no-external-field image Fig.~\ref{fig:interf2flows} (c). This layer corresponds to a steep plasma-density gradient, as evidenced by its visibility in the interferometric fringe patterns. The density jump has a characteristic scale of approximately 500 $\mu$m. 

At later times, 8--10~ns, two comparable density fronts emerge (Fig.~\ref{fig:interf2flows} (i, j)), indicating the outward propagation of this shock-like structure at an estimated speed of 75 km/s. These sharp density discontinuities are clearly visible in the interferometric fringe maps (Fig.~\ref{fig:interf2flows} (j)). Importantly, no such shock fronts were observed without external magnetic field, demonstrating the decisive influence of the externally applied transverse magnetic field on the plasma-jet interaction dynamics.

The external 14 T magnetic field corresponds to an Alfv\'en speed that is smaller, but comparable to the jet velocity for the measured plasma densities, $M_A \sim 1$, placing the interaction in a regime where magnetic pressure can become comparable to the ram pressure. An assessment was made for a wide range of parameters, as shown in Table \ref{Table_out}. Notably, the magnetic Reynolds number is large ($Re_M \sim 20$), implying that magnetic field diffusion across the flow is limited to thin boundary layers.


\section{\label{sec:Sym} Simulation}

\begin{figure*}
 \includegraphics[width=13 cm]{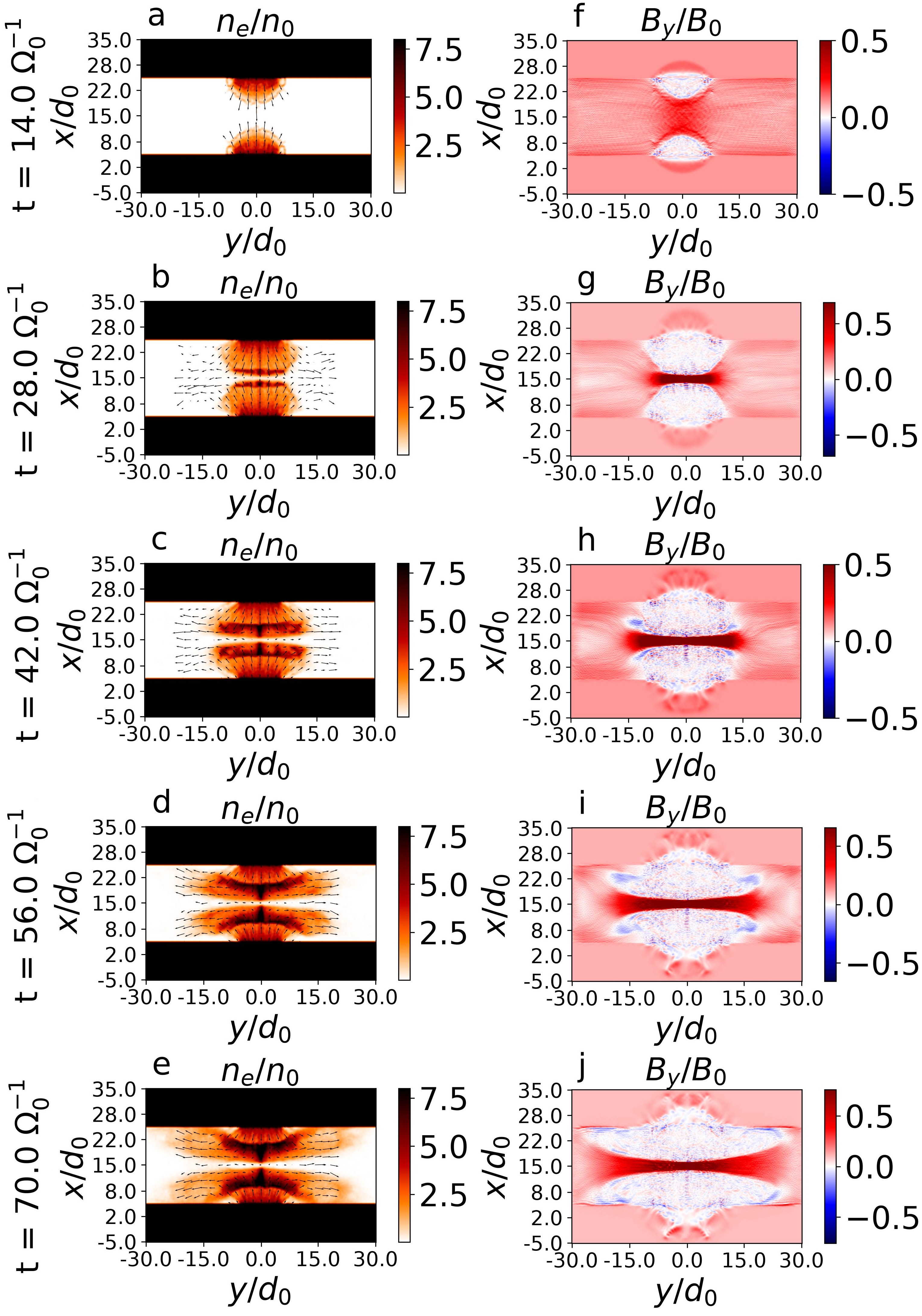}%
\caption{\label{fig:AKA_2d} Results of hybrid code AKA52 2D3V simulations for two counterstreaming plasma flows. Two-dimensional profiles of electron density of plasma (a, b, c, d, e)  and magnetic field $B_y$ projection (f, g, h, i, j)   at times 4$t_0$, 8$t_0$, 12$t_0$, 16$t_0$ and 20$t_0$, respectively. Arrows on the panels (a, b, c, d, e) indicate directions of ion motion, and arrow length demonstrates relative magnitude of ion velocity. Spatial scales are normalized to ion inertial length $d_0 = c/\omega_{pi} = 7.2$ $\mu$m, magnetic field is normalized to $B_0 \sim 140$ T, density is set to $n_0 = 10^{21}$ cm$^{-3}$, and time scales are normalized to the inverse ion cyclotron frequency  $t_0 = \Omega_{0}^{-1} = \frac{c*M_i}{q_i B_0} \sim 0.5$ ns.}
\end{figure*}

To explain the nature of the observed phenomenon, we present numerical simulations of the plasma using a hybrid (PIC - fluid) approach. We did not aim to perform a one-to-one simulation of the experiment due to computational resource limitations. However, our simulations are capable of reproducing important energy relationships (temperatures, kinetic energy) stored in the plasma and the self-generated magnetic field.

We use the hybrid code Arbitrary Kinetic Algorithm (AKA) \cite{akaGitHub} based on the common and well-accepted principles of the previous codes, such as Heckle \cite{Smets2011}, with advanced features such as an ablation operator and a six-component electron pressure tensor. In the numerical model, ions are described according to the Particle-In-Cell (PIC) formalism, while electrons are described by a ten-moment fluid model: density ($n_e$ equal to the total ion density due to neutrality), mass velocity ($V_{ex}$; $V_{ey}$; $V_{ez}$), and a six-component electron pressure tensor ($\widehat{P_e}_{xx}$; $\widehat{P_e}_{yy}$; $\widehat{P_e}_{zz}$; $\widehat{P_e}_{xy}$; $\widehat{P_e}_{yz}$; $\widehat{P_e}_{zx}$). Electromagnetic fields are considered in the low-frequency approximation (Darwin approximation), neglecting the displacement current. The generalized Ohm law contains three terms: (i) $\mathbf V_i \times \mathbf B$, where $\mathbf V_i$ is the mass velocity of ions; (ii) $(\mathbf J \times \mathbf B)/en$, the Hall effect, where $\mathbf J$ is the total current density that equals the rotor of $\mathbf B$, this term describes the decoupling of ions from the magnetic field; and (iii) $(\boldsymbol{\nabla} \cdot \mathbf P_e)/en$, the divergence of the electron pressure tensor, representing the contribution of the electron fluid \cite{sladkov2021}. Ion collisions are accounted for using the Takizuka-Abe binary collision model \citep{takizuka1977}.

The magnetic field and density in the model are normalized to $B_0 = 140$ T and $n_0 = 10^{21}$ cm$^{-3}$, respectively, from which follows the normalization of the other quantities. The density determines the ion inertial length, $d_{0}:= d_{i0}  \sim 7.2 \, \mu m$, which sets the length scale. The box has a 2D rectangular shape with dimensions $(40d_0 \times 60 d_0)$. The magnetic field determines the time normalization, as the time is normalized to the inverse ion cyclotron frequency, $ t_0 = \Omega_0^{-1}$, which is approximately $0.5$ ns. The velocities are normalized to the Alfvén speed $V_0 \sim 96 $ km/s calculated from the normalized quantities $B_0$ and $n_0$. In the simulation, the target density was initialized and maintained at a level of $n_0$ (i.e. $10^{21}$ cm$^{-3}$). The pixel size is chosen so as to resolve the ion inertial length in the interaction region.

Continuous plasma production resulting from the laser-target interaction is simulated using an ablation operator \cite{sladkov2020}, which accounts for the heating of ions and electrons. The operator of particle production maintains constant target density, simulating a reservoir for a target with solid-state density. This approach is adequate, as we observe in the experiment that after irradiation the target serves as a prolonged (several tens of ns) source of plasma. The heating operator maintains a constant electron pressure in a thin layer near the target surface, creating a pressure gradient along the normal to the target surface; this generates an electric field and accelerates ions, giving rise to plasma expansion (see Fig. \ref{fig:AKA_2d} (a)). The diameter of the heated area is 15$d_0 \sim$ 100 $\mu$m.

The evolution of a single plasma flow and of two counterstreaming plasma flows in vacuum was modeled in our previous work \cite{zemskov2026magnetic}. The simulation results (Figures 5, 6 in Ref. \cite{zemskov2026magnetic}) demonstrate that the AKA code reproduces the experimentally observed flow topology quite well.

Figures \ref{fig:AKA_2d} illustrate the evolution of the interaction between two counterstreaming plasma flows in vacuum with an external uniform magnetic field $B_{y,ext} = 0.1 \cdot B_0 = 14$~T. The electron pumping pressure in the target was set to $P_{e, pump} = P_0$, corresponding to a temperature  $T_{e, pump} \sim 48$ eV. The black arrows on the panels (a, b, c, d, e) indicate directions of ion motion and the arrow length demonstrates a relative magnitude of ion velocity. Already at the first presented stage $t = 14 t_0$, the plasma expands and pushes out the external magnetic field (Fig. \ref{fig:AKA_2d} (f)). As a result, the initially uniform magnetic field becomes substantially modified. When the plasma flows approach each other ($t \gtrsim 28 t_0$), the magnetic field in the center of the simulation domain is compressed into a thin layer with a characteristic width $\delta l \sim 4 d_0$ (Fig. \ref{fig:AKA_2d} (g, h)). The high ram pressure of the plasma plumes, compressing this plasma-magnetic layer from both sides, leads to amplification of the magnetic field up to $B_y \sim 0.5 B_0$, which is a factor of 5 larger than the initial field. Thus, the magnetic pressure in this layer becomes comparable to the ram pressure of the two incoming plasma flows, resulting in the redirection of plasma flows to the y-direction. Indeed, as shown by the arrows in Fig. \ref{fig:AKA_2d} (d, e), the ions change their directions and move along the magnetic layer.

To verify the validity of the effects observed in the 2D simulations, additional three-dimensional modeling was performed using the AKA52 code. In general, the simulation parameters, such as the target separation, the initial plasma temperature and density, and the amplitude of the external magnetic field were the same as those used in the two-dimensional calculations. However, to reduce computational costs, the spatial resolution was decreased, although not critically. The simulation box had a rectangular shape with dimensions $(40d_0 \times 60d_0 \times 40d_0)$ and the resolution $d_0/4$. In addition, to speed up the calculation, the pump pressure was increased by a factor of 20 compared to the 2D case, $P_{e, pump} = 20 P_0$, which corresponds to the temperature $T_{e, pump} =$  960 eV. To maintain a balance between the energy of the hot electrons excited by the laser in the target and the energy of the magnetic field, the duration of excitation of the hot electrons was reduced by 10 times, $\sim$ 5 $\Omega_{0}^{-1}$. 

\begin{figure*}
 \includegraphics[width=14 cm]{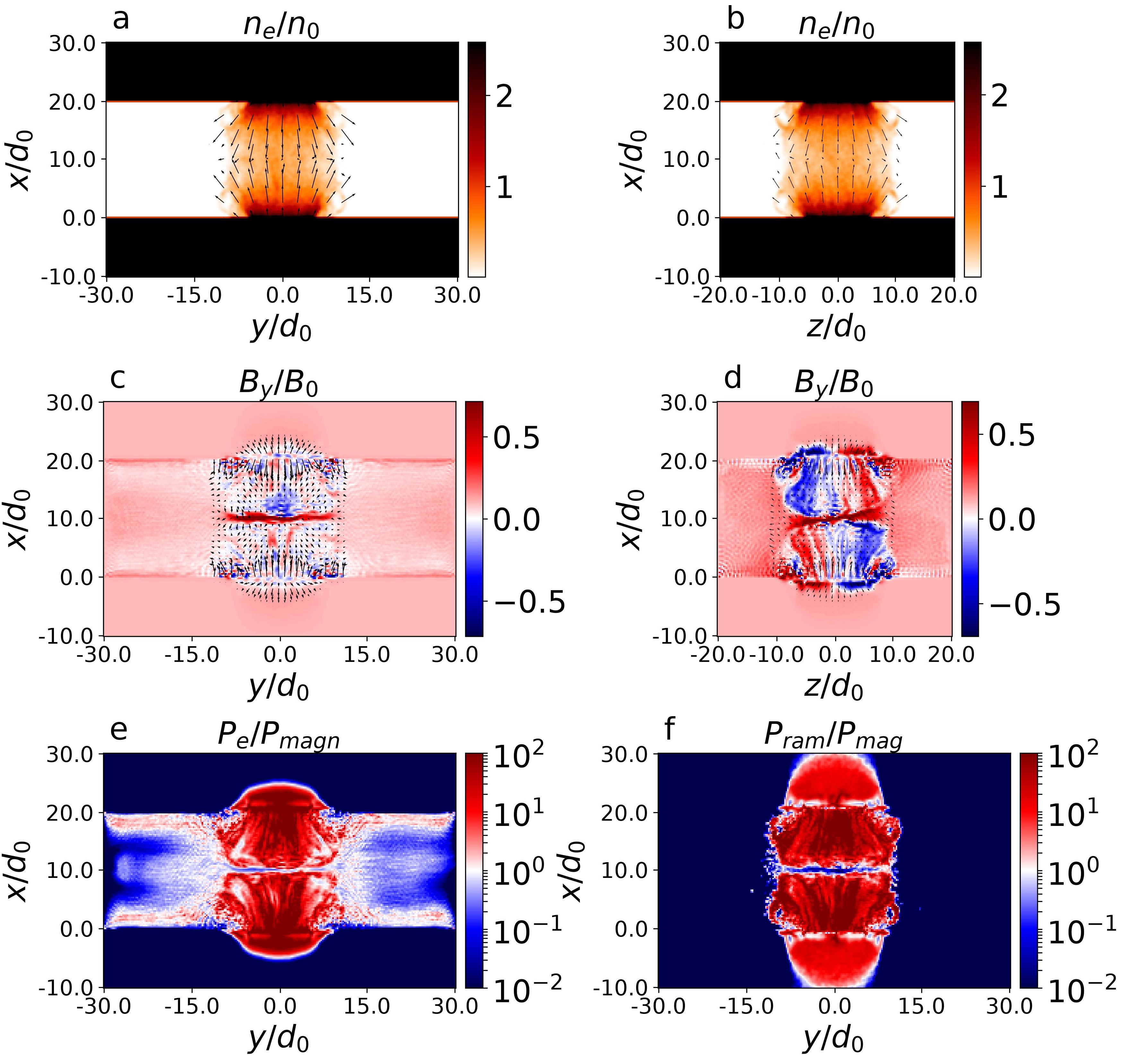}%
\caption{\label{fig:AKA_3d} Results of hybrid code AKA52 3D3V simulations for two counterstreaming plasma flows at $t = 8t_0 = 8\Omega_0^{-1}$. Two-dimensional central ($y = 0$ or $z = 0$) slices of electron density of plasma (a, b) and magnetic field $B_y$ projection (c, d). Black arrows on the density panels (a, b) indicate directions of ion motion, and arrow length demonstrates relative magnitude of ion velocity. Black arrows on the magnetic field panels (c, d) indicate electric field $\boldsymbol{E} = -[\boldsymbol{V_e}\times \boldsymbol{B}] - \frac{1}{en_e}\nabla_. \boldsymbol{P_e}$ (the Hall effect included), the arrow length demonstrates relative magnitude. Spatial scales are normalized to ion inertial length $d_0 = c/\omega_{pi} = 7.2$ $\mu$m, magnetic field is normalized to $B_0 \sim 140$ T, density is set to $n_0 = 10^{21}$ cm$^{-3}$, and time scales are normalized to the inverse ion cyclotron frequency  $t_0 = \Omega_{0}^{-1} = \frac{c*M_i}{q_i B_0} \sim 0.5$ ns.}
\end{figure*}

The fully three-dimensional modeling confirmed the main observations of the 2D simulations. However, owing to its computational complexity, the 3D modeling did not allow us to follow the long-term dynamics for times $t > 10\,\Omega_0^{-1}$. Figure~\ref{fig:AKA_3d} presents the 3D simulation results obtained at $t = 8\,\Omega_0^{-1} \sim$ 4 ns. The corresponding stage in the 2D simulation is shown in Fig.~\ref{fig:AKA_2d}(b - c). Figure~\ref{fig:AKA_3d} displays two-dimensional slices of the plasma density at $z = 0$ (panel a) and $y = 0$ (panel b). The black arrows in these panels indicate the ion velocity directions. It is clearly seen that in the central collision region of the counterstreaming flows, the process of particle redirection begins, similar to that observed in the 2D simulation (Fig.~\ref{fig:AKA_2d}).

Figures ~\ref{fig:AKA_3d} (b) and (c) show slices of the magnetic field component $B_y$ at $z = 0$ and $y = 0$, respectively. As in the 2D case (Fig.~\ref{fig:AKA_2d}(g)), a layer with a magnetic field amplified by more than a factor of $\sim 5$ is formed. Its thickness, $\sim 2\,d_0$, is somewhat smaller than in the 2D simulation, that is easily explained by the higher initial pressure in this calculation, $P_{e, pump}^{3D} = 20 P_{e, pump}^{2D}$. Taking into account the local ion density in the central region, $n_i = n_e/Z \sim n_0/4$, the thickness of this layer matches the local ion inertial length $d_i = c/\omega_{pi}$. It is worth noting a slight inclination of the magnetic layer in the $y = 0$ plane (Fig.~\ref{fig:AKA_3d}(d)), which is caused by the contribution of the self-generated Biermann battery magnetic field \cite{biermann1950ursprung}. The contribution of this field to the $B_y$ component could not be accounted for in the 2D simulation due to the inherently three-dimensional geometry of the Biermann field. However, as shown by the 3D simulation, this self-generated field does not exceed the amplified external field, so its influence on the plasma dynamics is not substantial. Additionally, Fig.~\ref{fig:AKA_3d}(c) reveals modulations of the magnetic layer surface that were not previously observed in the 2D simulation. The black arrows in Fig.~\ref{fig:AKA_3d}(c, d) depict the direction of the local electric field 
$$\boldsymbol{E} = -[\boldsymbol{V_e}\times \boldsymbol{B}] - \frac{1}{en_e}\nabla \cdot \boldsymbol{P_e}$$
lying in the plane of the figure, i.e. $(E_x, E_y)$ in panel (c) and $(E_x, E_z)$ in panel (d). Importantly, in Fig.~\ref{fig:AKA_3d}(c) the electric field vectors on the two sides of the magnetic layer point in opposite directions, which produce opposite $[\boldsymbol{E} \times \boldsymbol{B}]$ particle drifts along the $z$-direction. Note that in Fig.~\ref{fig:AKA_3d}(d), at the boundaries of the magnetic layer ($z = \pm 12 d_0$, $x = 10 d_0$), the electric field is directed so as to cause particle drift out of the magnetic layer region.

Figure~\ref{fig:AKA_3d}(e) shows a slice ($z = 0$) of the ratio of the electron pressure $P_e = \frac{1}{3} (P_{e,xx} + P_{e, yy} + P_{e, zz})$ to the magnetic pressure $P_{mag} = 0.5 B^2$. One can see that the electron pressure exceeds the magnetic pressure throughout the entire region occupied by the plasma, except for a thin layer with an amplified magnetic field. Within this thin layer of thickness $\sim d_i$, the magnetic pressure significantly exceeds the electron pressure. A similar picture is observed for the ram pressure $P_{ram} = 0.5 \rho V^2$ (Fig.~\ref{fig:AKA_3d}(f)), which is dominated by high-velocity heavy ions. Estimates based on the experimental parameters and modeling (not shown) indicate that at late times ($t > 8\,\Omega_0^{-1} \sim 4$ ns), the ram pressure exceeds the electron pressure; therefore, at these times it is the ions that govern the large-scale plasma dynamics. Despite the high ion velocities, we see that in the region of the amplified magnetic field layer, $P_{ram} < P_{mag}$. Using the relation $\frac{P_{ram}}{P_{mag}} = \frac{V^2}{V_A^2} = \frac{R_{Li}^2}{d_i^2}$, it is demonstrated that, inside the magnetic layer, the ion cyclotron radius $R_{Li}$ is smaller than the ion inertial length $d_i$, which characterizes the thickness of this layer. Thus, upon approaching the vicinity of the magnetic layer, ions are efficiently redirected by the strong magnetic field, as observed in both the experiment and the 2D and 3D simulations. Further compression and amplification of the magnetic field is also energetically unfavorable, since the magnetic pressure already exceeds the ram pressure. All this leads to large-scale plasma deceleration, subsequent redirection along the layer, and the formation of sharp plasma density jumps, as demonstrated in the experiment and 2D modeling.

\section{\label{sec:Discus}Discussion and Conclusions}

In this work, we have presented experimental and numerical investigation of the interaction of two counterstreaming laser-produced plasma jets in the presence of a uniform transverse magnetic field. The experiments, performed at the PEARL facility using femtosecond laser pulses, demonstrate that the application of a 14 T external field fundamentally alters the dynamics of the colliding flows compared to the previously reported unmagnetized case \cite{zemskov2026magnetic}. While the baseline configuration without external field exhibits stagnation and lateral redirection driven by the compression of self-generated toroidal magnetic fields, the magnetized case reveals a distinct regime characterized by the formation of sharp density discontinuities and narrow magnetic-field layers.

The central experimental observation is the appearance of a thin, quasi-stationary density jump in the interaction region, which develops within a few nanoseconds after the jet collision and subsequently propagates outward at a speed of 75 km/s. This feature, clearly visible in interferometric, shadowgraphic, and density maps, has no counterpart in the unmagnetized experiments and can be directly attributed to the presence of the externally applied transverse magnetic field. 

Hybrid PIC--fluid simulations performed with the AKA code reproduce the principal features of the experiment with satisfactory agreement. The simulations show that the initially uniform 14 T field is advected and compressed into a thin layer as the two plasma jets approach each other. Within this layer, the magnetic field strength is amplified by a factor of approximately 5, reaching values at which the magnetic pressure becomes comparable to the ram pressure of the incoming flows. This pressure balance naturally explains both the formation of the observed density jump and the subsequent redirection of the plasma along the magnetic field lines. The qualitative agreement between the experimental observations and the numerical results supports the interpretation that the compression of the external magnetic field by the converging plasma ram pressure is the dominant mechanism responsible for the observed phenomena.



From an astrophysical perspective, the compression-driven amplification mechanism explored here offers a promising complementary pathway to magnetic-field amplification in interstellar media. 
The present results demonstrate that a simple compressive pile-up of an initially weak ambient field by converging flows can provide an efficient, non-turbulent channel for magnetic-field amplification. This mechanism operates on timescales much shorter than those required for turbulence to develop and saturate. On the other hand, a large-scale magnetic field which is parallel to the shock front can stabilize the front and suppress the growth of the turbulence.
Observations in the X-ray and radio ranges reveal filaments with a length of about a few $\times10$~pc and a magnetic field of up to $100\; \mu G$ can be formed by strong shocks with efficient cooling or multiple shock compression episodes. Notably, such filaments allows us to probe efficiency of the particle acceleration, which can be as high as $\eta_{acc}\sim3/4$  \cite{2027JHEAp..5500710N}.


In conclusion, the present experiments establish a new laboratory platform for studying magnetized plasma interactions in a regime where the external magnetic field is dynamically compressed to strengths comparable to the plasma ram pressure. The observed sharp density discontinuities and the amplification of the transverse magnetic field are well explained by a simple physical picture of advection and compression, supported by hybrid simulations. The results demonstrate that compressive amplification of ambient magnetic fields by converging plasma flows is an efficient, non-turbulent mechanism that may contribute to magnetic-field amplification in astrophysical shocks, complementing the more commonly discussed turbulent and instability-driven processes. 

\begin{acknowledgments}

The experiments at the PEARL laser facility were supported by the Russian Science Foundation, project No. 24-62-00032. The numerical simulations were carried out with the support of the 10th project of the National Center for Physics and Mathematics (NCPhM) ``Experimental laboratory astrophysics and geophysics''.

\end{acknowledgments}

\bibliographystyle{apsrev4-1}
\bibliography{ShockBIB}

\end{document}